\documentclass[runningheads]{llncs}
\usepackage{graphicx,hyperref,subfigure}
\usepackage{multirow,comment}
\usepackage[table]{xcolor}
\usepackage[belowskip=-20pt,aboveskip=-5pt, font=small, labelfont=bf]{caption}

\begin{document}
\title{Necknasium\textsuperscript{TM}: A Virtual Reality Rehabilitation Game for Managing Faulty Neck Posture\thanks{This study has been approved by the IRB Approval Committee at Faculty of Physical Therapy, Cairo University (P.T.REC/012/004171).}}
\author{Aliaa Rehan Youssef\inst{1}\orcidID{0000-0002-5345-5936}\and
Mohammed Gumaa\inst{2}\orcidID{0000-0003-1658-0129}\and
Ahmad Al-Kabbany \inst{3,4} \orcidID{0000-0002-3941-0061}}
\authorrunning{Rehan Youssef et al.}
\institute{Faculties of Physical Therapy and Engineering, Cairo University, Egypt
\email{aliaa.rihan@pt.cu.edu.eg}
\and
Trust Research Center, Mokkatam, Cairo, Egypt
\email{m.gumaa.bakry@gmail.com}
\and
Multimedia Processing, Communication, and Interaction Lab, Arab Academy for Science, Technology, and Maritime Transport, Alexandria 21937, Egypt
\email{alkabbany@ieee.org,alkabbany@aast.edu}\\
%\url{https://www.linkedin.com/company/intelligent-systems-lab} 
\and
VRapeutic Inc., Ottawa K2M1T2, Canada}

\maketitle              
% typeset the header of the contribution
%
\begin{abstract}
This study is concerned with the application of virtual reality (VR) in the rehabilitation programs for faulty neck posture which is a primary source of neck pain (NP). The latter is a highly prevalent musculoskeletal disorder that is associated with serious societal and economic burden. VR has been shown effective in the physical rehabilitation of various diseases. Specifically, it has shown to improve patients’ adherence and engagement to carry out physical exercises on a regular basis. Many games have been used to manage NP with different immersion levels. Towards this goal, we present a VR-based system that targets a specific neck problem,  the so called forward head posture (FHP), which is a faulty head position that abnormally stresses neck structures. The system can also generalize well to other neck-related disorders and rehabilitation goals. We show the steps for designing and developing the system, and we highlight the aspects of interaction between usability and various game elements. Using a three-point scale for user experience, we also present preliminary insights on the evaluation of the system prototype, and we discuss future enhancement directions based on the users’ feedback.\footnote{All the authors contributed equally to this research}

\end{abstract}

\section{Introduction}
With the increased use of handheld devices, prolonged office work hours, and other activities that force neck into bad posture, neck pain (NP) is increasing as a global problem nowadays. There are other more common musculoskeletal disorders, yet NP is not getting less significant \cite{ariens2001psychosocial,cohen2015epidemiology}. Several studies have highlighted the severe societal and economic burden of spinal pain especially the neck in the US \cite{dieleman2020us}, other African \cite{rushton2019neck} and Scandinavian \cite{safiri2020global} countries. This just further emphasizes the highly-prevalent and global nature of this problem and its growing adverse impact as a burden on the healthcare systems and the world’s economy.

The research literature on applying technology in physical therapy and rehabilitation is immense. Diverse forms of software and hardware technology can be found in the literature including 2D and 3D serious games \cite{martins2020physioland}, tablets and VR headsets \cite{kim2019nexercisevr}, sensors, therapeutic robots \cite{gonzalez2021robotic}, and tele-rehabilitation-based delivery \cite{seron2021effectiveness}. %Also, the growing incorporation of technology in rehabilitation programs has covered multiple contexts including screening, assessment, diagnosis \cite{maceira2019wearable}, and management/treatment \cite{feng2019virtual}.

%The research literature on applying technology in physical therapy and rehabilitation is immense. Diverse forms of software and hardware technology can be found in the literature including 2D and 3D serious games \cite{martins2020physioland}, tablets and VR headsets \cite{kim2019nexercisevr}, sensors, therapeutic robots \cite{gonzalez2021robotic}, and tele-rehabilitation-based delivery \cite{seron2021effectiveness}. Also, the growing incorporation of technology in rehabilitation programs has covered multiple contexts including screening, assessment, diagnosis \cite{maceira2019wearable}, and management/treatment \cite{feng2019virtual}.

Technology-based therapy for neck disorders and pain has been the focus of an ever-growing body of research \cite{salim2019quantifying}. There are three main directions for incorporating technology in rehabilitation programs. First, it has been used as the medium through which specific exercises are being displayed to the user, which capitalizes on the use of technology as a source of motivation for completing the exercises \cite{kim2019nexercisevr}. Second, technology was incorporated in other studies as means for tracking motions \cite{salinas2021camera}. This would represent a repository of performance data acquisition according to which therapy plans can be developed and modified. Third, technology is some-times used as means of tele-rehabilitation \cite{crosetti2021impact}. The research on this application, in particular, had thrived while COVID-19 was soaring worldwide.

Previous research referred to the role of deep neck flexor (DNF) muscles strength and endurance as principal factors for improving NP especially in patients’ with FHP \cite{sikka2020effects} which is strongly associated with NP in adults \cite{iqbal2021efficacy}. However, to authors’ knowledge, no VR games was specifically designed to target faulty head posture or FHP in those patients. Accordingly, developing a VR-based deep cervical training might be useful in improving FHP, decreasing NP, and improving users’ acceptance to exercises. In this research, we present the design, development, and prototype evaluation processes of a new VR-based module named Necknasium\textsuperscript{TM} for correcting FHP. This module takes place in a gymnasium (hence the name) where the users are instructed to do specific repetitions of movements that involves lifting a weight for a specific vertical distance in order to help improving their FHP and hence manage a potential primary source of their NP. %We start by highlighting the details of the system design process where we cover different game design aspects such as game leveling in the light of the intended training outcomes. We also discuss challenges related to capturing and quantifying neck movements using motion sensors as a part of the system design process. Afterwards, we explain the details of the VR software development process and the user’s experience aspects considered during this process. This is followed by developing and implementing an approach for conducting a preliminary prototype evaluation. This covers the usability and the user-experience aspects of the game. 
The contributions of this research can be summarized as follows:
\begin{enumerate}
    \item We specify a set of intended training goals for the trainee and the therapist and we lay the design foundations based on these goals.
    \item We highlight a set of challenges that we faced during the implementation of the system design framework, and we present the adopted alternatives as well as the pros and cons of the initial and eventual approaches. 
    \item We present the software development framework adopted to realize the pro-posed system and the different game elements that constitute this framework.
    \item We present a framework for preliminary prototype evaluation and its implementation on three healthy asymptomatic users; two undergraduate engineering students and a junior VR developer. 
    \item We suggest directions for future development based on the implementation of the proposed framework.
\end{enumerate}
With just three healthy participants, this research is meant to give preliminary insights on the feasibility and the potential of the proposed VR-based training for NP.

The rest of this article is structured as follows. In section 2, we highlight previous research on gamification-based neck rehabilitation that is relevant to the proposed study. In section 3, we present the design and development of the proposed system. Section 4 features the results obtained from the preliminary prototype evaluation, before the study is concluded in section 5.

\section{Related Work}
%In this section, we cover previous research endeavors on using technology for NP management in more detail. Although the literature review will cover all technology-based approaches in general, a special focus will be given to VR-based solutions for their relevance with the system being proposed in this study. For most of the highlighted approaches, we will mention the rehabilitation goals and the results as well.

The authors of \cite{kim2019nexercisevr} proposed a smartphone-based VR exergame app for deep neck flexor (DNF) endurance exercise. Mainly, it addressed posture correction and improving the range of motion (ROM). The prototype validation was accomplished on three healthy individuals. However, this system still lacks scientific evidence of its clinical efficacy. Another system was proposed in \cite{mihajlovic2018system}. The user is asked to chase a butterfly. This task was chosen to train head-neck movement coordination during high speed motion in addition to static positioning when the butterfly standstill in the air. The system consisted of assessment and treatment modules that are consumed through Oculus Rift\textsuperscript{TM}. Lastly, the research was concluded with a case study that featured higher interest and enjoyment.

%The authors of \cite{bahat2015cervical,bahat2020predictors}

The authors of \cite{bahat2020predictors} addressed the ROM in addition to the movement velocity and accuracy. They developed a system for clinic and home-based neck rehabilitation that featured two main components. First, a hardware component with an Oculus Rift\textsuperscript{TM} and 3D motion sensors. Those sensors include a gyroscope, an accelerometer, and a magnetometer as well as Complementary Metal-Oxide-Semiconductor (CMOS) sensor for positional tracking. Second, there is a software component which is represented by the VR modules and the motion data analysis programs specially designed for that study. Three modules were developed, each of which addresses ROM, velocity, and accuracy. The modules were designed to visually-stimulate cervical motion by the patient. This study supported VR efficacy in the assessment and treatment as evident by improving pain, disability, movement velocity and accuracy, and quality of life after up to three months follow-up.

Towards the reduction of pain and functional disability, VR was shown to have an advantage over traditional exercises in \cite{rezaei2019novel}. This study mainly targeted proprioception which refers to the sense of movement and position of the neck. The proposed Cervigame\textsuperscript{\textregistered} is a video game in which the user is required to control the movements of a rabbit using their head movements. The amount of required control progresses over fifty levels. The proposed systems involved placing a reflective marker just above the eyebrows of the patient. Moreover, a Head Mouse Extreme\textsuperscript{\textregistered} (Origin Instruments Corporation, Grand Prairie, Texas, USA) was position at the top of laptop’s monitor and aimed on reflective markers. This setup enabled head motions to control a pointer movement that was placed on the laptop’s monitor.

%The authors of \cite{morales2020effects} reported similar effectiveness for immersive VR and traditional exercises with regards to decreasing pain and reducing disability. How-ever, VR was shown to be more effective in reducing pain-related fear of movement. Their system represented an immersive smartphone-based app. This system featured a VR Vox Play glasses with a head-mounted display (HMD) clamping system to which the smartphone (LG Q6) was attached. The associated software comprised two VR applications, namely, "Full-dive VR" in which the user is required to tilt their neck only, and "VR Ocean Aquarium 3D" in which the user is required to do flexion, extension and rotation movements in addition to a sensory element represented by the sound of the sea.

It is worth noting that none of the described studies targeted neck retraction, which is a translation movement performed specifically by the DNF muscles and is the primarily exercise employed to correct FHP.

\section{Proposed Methods}
%In this section, we start off by highlighting the principal requirements which had guided the software design process. We start by listing those requirements. Then, we highlight our system design experience with regards to motion tracking. Moreover, we elaborate on the system perspectives based on physical therapists' and the trainees’ feedback. Lastly, we cover some aspects related to the applicability of the system in FHP and other neck disorders.

\subsection{Identifying the Design Requirements}
The system design requirements are inspired by the following aspects: 1) the targeted user experience from the perspective of the rehabilitation professionals and trainees, 2) the targeted exercise dynamics for FHP correction, 3) and the targeted level of training customization or personalization. We require the proposed system to satisfy the following points:
\begin{enumerate}
    \item The system should provide the user to set different ranges of motion during neck retraction (backwards translation movement of the whole head and neck).
    
    \item The system should be capable of doing automatic calibration in order to accommodate different ranges of motion during neck retraction, especially when the exercise is performed remotely.
    
    \item The task to be accomplished within the virtual environment, which represents the neck exercise, that should look as realistic as possible and should be a real-life task to guarantee an acceptable level of motivation and engagement.
    
    \item The designed exercise should target both strength (ability to produce muscle force) and endurance (ability to hold the target position as long as possible while exerting low sustained force) of muscles.
\end{enumerate}

\subsection{Motion Tracking and Sensing Challenges}
Akin to \cite{han2019novel}, we initially tried to design and develop external inertial motion units (IMUs) which were supposed to communicate with the exercise on the VR headset through a WiFi module. The main role of the IMUs was to sense the motion of the neck in real-time, send the magnitude of this motion, and then this motion is translated to a game event in the VR environment. The sole purpose of building our own IMUs was to have full control over the type of data they collect throughout the exercise. However, the manufactured IMUs had suffered from sporadic noise and unstable behavior often. This was quite apparent when the readings of the IMUs were compared to motion video analysis using the open source Kinovea software\textsuperscript{\textregistered}\footnote{https://www.kinovea.org/ }. Accordingly, we resorted to the motion sensors embedded in the VR headset.

An Oculus Quest 1\textsuperscript{\textregistered} VR headset was used throughout the software development and the implementation of this study. This standalone VR headset offers access to several types of motion information. Figure 1(a) shows the types of motion information which we had access to from the VR headset being shown on the left of the main menu of the system. Following several experiments, we found that the device position can be relied on to detect the neck retraction movement, which is required during the FHP correction exercise. We also found that the angular velocity or acceleration can be used to detect erroneous movements during the exercise. Exceeding a threshold of angular velocity or acceleration is set to display a warning to the user.

\begin{figure}
\begin{center}
\subfigure[]{
\frame{\includegraphics[height=1in]{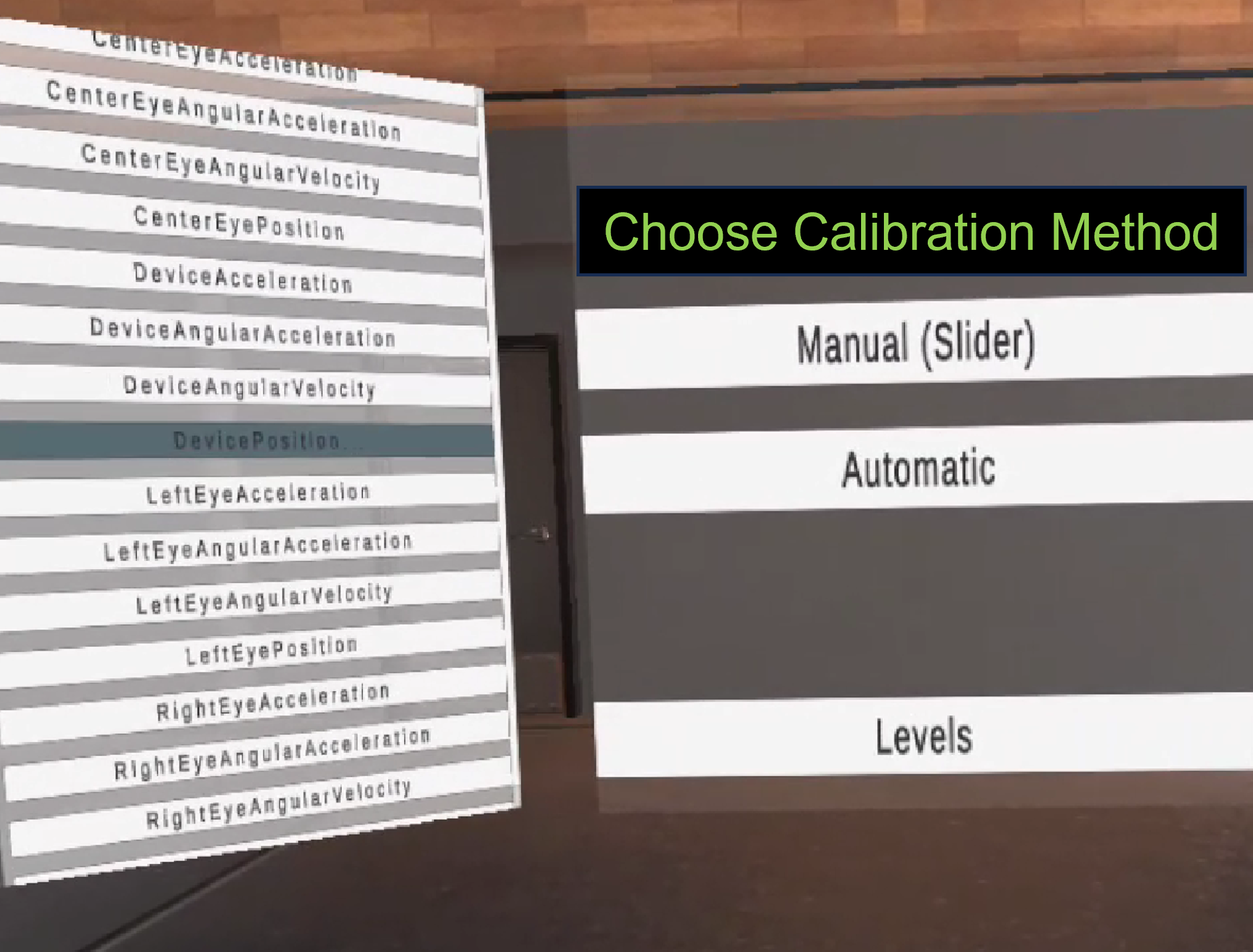}}}
\subfigure[]{
\frame{\includegraphics[height=1in]{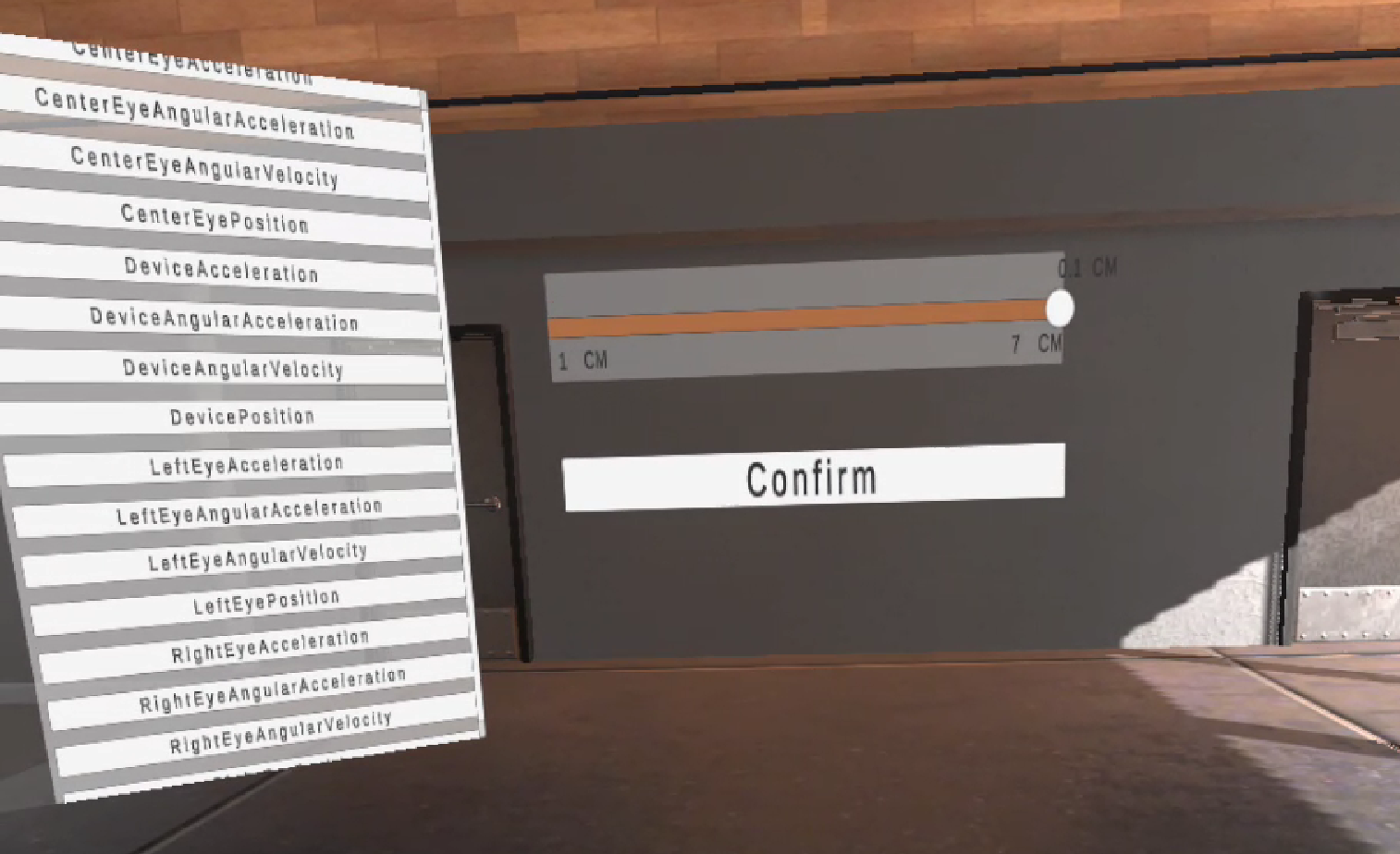}}}
\subfigure[]{
\frame{\includegraphics[height=1in]{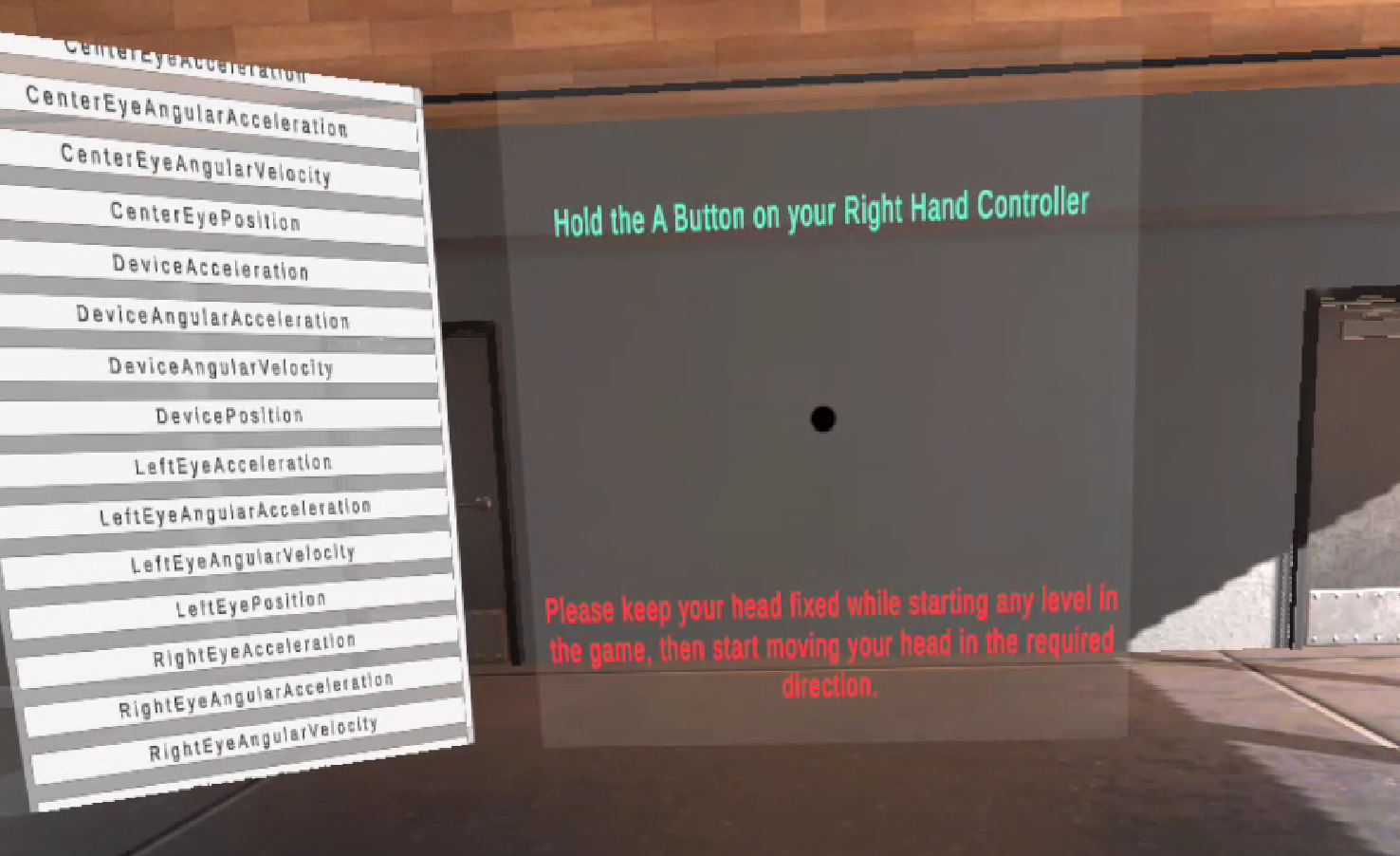}}}

\subfigure[]{
\frame{\includegraphics[height=1.2in]{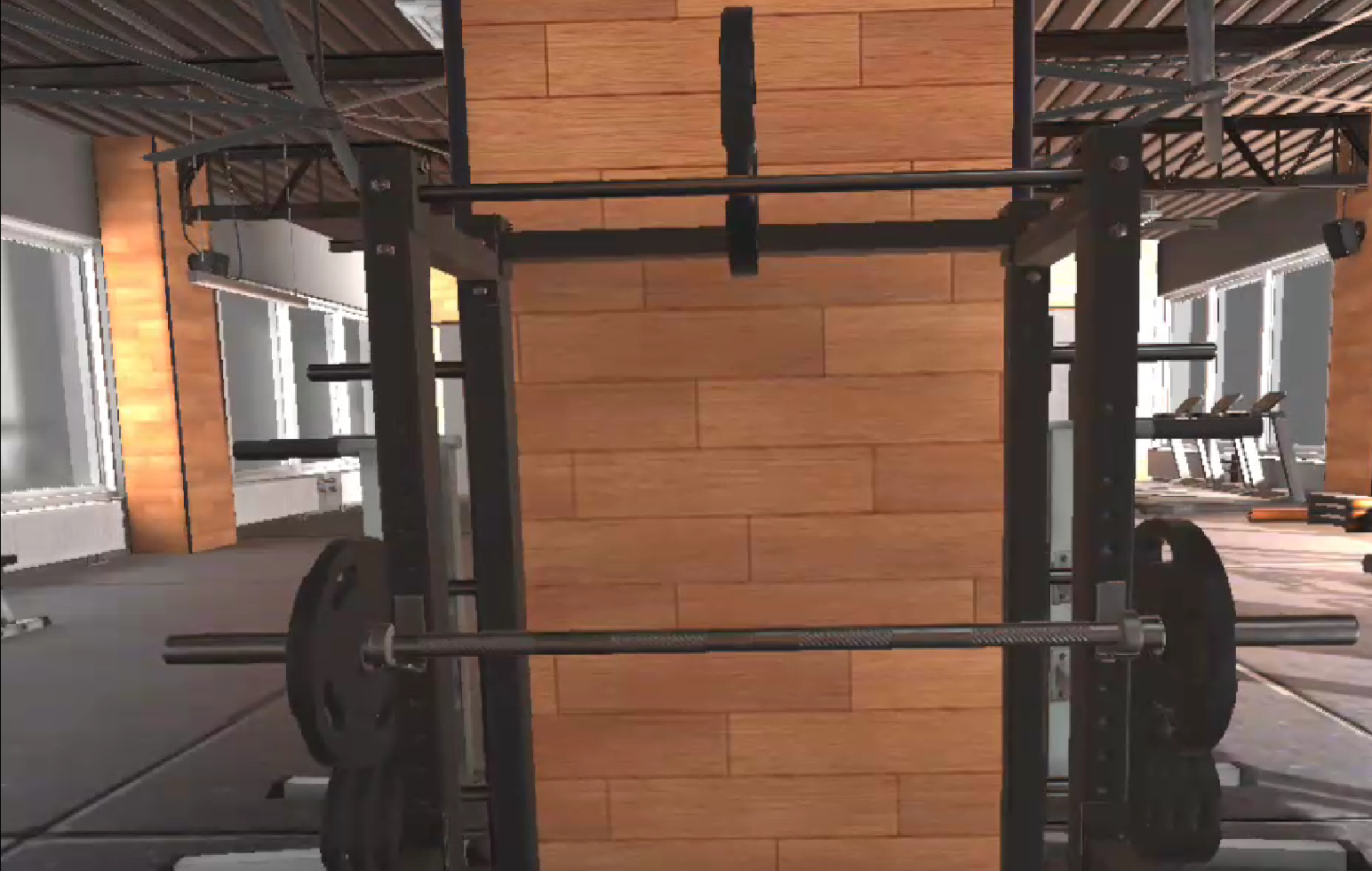}}}
\subfigure[]{
\frame{\includegraphics[height=1.2in]{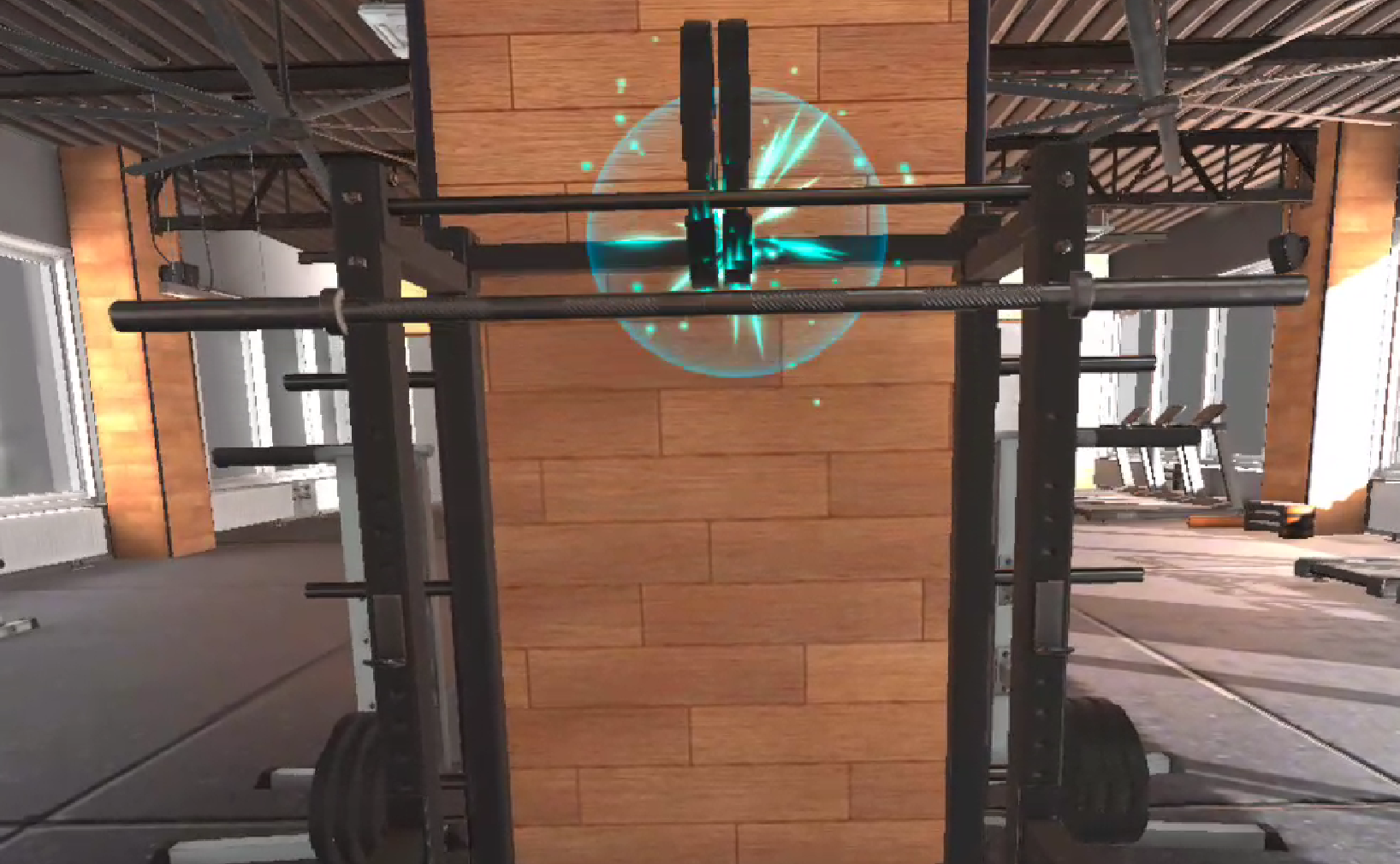}}}
\caption{Screenshots from inside the prposed system showing a) the types of motion info that can be acquired from Oculus, b) manual calibration and c) automatic calibration menus, and two screenshots (d and e) from inside the gymnasium VR environment showing the bar to be lifted at its initial and final position respectively.}
\label{Fig:OcParam}
\end{center}
\end{figure}

%Advantages and disadvantages of both approaches

\subsection{System Perspectives}
As shown in the main menu in Fig. 1(a), automatic and manual calibrations are supported by the system. From the therapist point of view, it is helpful to be able to set the maximum range of neck retraction manually to ensure personalized rehabilitation based on each individual patient’s needs. On the other hand, in remote rehabilitation scenarios, it might also be helpful to make the system capable of doing an automatic calibration. This might serve as a validation of the subjective distance that a therapist can estimate remotely. When the manual calibration is chosen, the menu in Fig. 1(b) is displayed, where the user can set a maximum retraction distance using a slider. The automatic calibration, however, requires the user to long-press a button on the controller twice, at the beginning and the end of the neck retraction motion, then the maximum retraction distance is saved automatically.

The task inside the VR environment is to raise a weight bar in a gym, hence the name–Necknasium, for a certain number of repetitions. Following the calibration, step, from a user’s/trainee’s perspective, the system provides six levels for doing the aforementioned task. The first three levels require the user to perform a neck retraction movement to at least 30\%, 60\%, and 90\% of the maximum range, respectively. Each of these levels require thirty retractions or repetitions in order to train DCF muscles strength. Levels four, five, and six require the user to do the same task as the first three levels except that they also require the user to stay in the retraction position for a longer time to target the neck muscles’ endurance. Fig. 1(d,e) show two screen-shots from inside the VR environment where the bar is at its initial and end positions, respectively, where the end position is associated/marked with a visual effect and an auditory stimuli to complement the VR environment.

\subsection{Addressing Various Scenarios}
It is worth mentioning that the proposed system, with its reliance on the em-bedded sensors in the VR headset, can be adopted in other use cases for neck disorders, not just FHP-related NP that focuses on improving retraction movement. Because we have access to all the information about device position, velocity, and acceleration in the three dimensions as a function in time, the proposed system can be incorporated in rehabilitation plans that involve other movements other in different plans such as neck bending and extension. 

\section{Results and Discussion--A Preliminary Prototype Evaluation}
The clinical effectiveness of the proposed system is beyond the scope of this study. Yet previous studies that highlighted the influence of neck exercises regarding reduction of pain and its associated disability. These studies include [4,5] and [12]. Instead of clinical effectiveness, we aim at investigating the feasibility of realizing engaging exercises through VR exergaming, which would positively impacts patients’ adherence to exercises and hence improving sustainability of physical therapy rehabilitation to achieve its target outcome.

As stated in the previous section, neck movements is tracked in real-time using the motion sensors embedded in the VR headset. Hence, those senors represent a source of quantitative evaluation for the user performance and progress achieved throughout the therapy program. In addition, the system provides real-time feedback that motivates patients’ to exert maximum possible effort during the training process. Based on this continuous real-time monitoring, a therapist would develop and modify the rehabilitation program to target changing individual’s needs. These data include basic statistics (e.g., maximum, minimum, range) of the distance moved by the neck throughout the repetitions of the exercise.

Towards conducting a preliminary prototype evaluation, the users are instructed to complete the following settings:
\begin{enumerate}
    \item Level 3 of the VR module, which is the last level that addresses the DNF muscles’ strength. In this level, the user is required to complete thirty neck movements, each of which should last for at least six seconds. We refer to this setting as Setting 1 in the rest of this section.
    \item Level 6 of the VR module, which is the last level that addresses DNF muscles’ endurance. In this level, the user is required to complete thirty neck movements, each of which should last for at least ten seconds. We refer to this setting as Setting 2 in the rest of this section.
\end{enumerate}

Three volunteers were recruited to participate in this evaluation, all of them are males in the age range from 21-24 years old. Two of those volunteers are undergraduate engineering students with one year of experience in game development. The third participant is a junior game developer with one year of work experience in VR development. All participants were healthy and asymptomatic with no history of NP. Initially, all participants received full description of the study purpose and the required tasks. Prior to actual game trials, they were shown a video\footnote{https://youtu.be/6C-wfV27bzI} that explains how traditional neck retraction exercise looks like. As will be shown below, this part becomes a core component of this participation in the rest of the evaluation study.

Two aspects of evaluation were covered in this study, namely, a subjective rating for engagement and preference of VR-based intervention and another subjective rating for user’s experience. The experience was evaluated on a scale from -/+1. The former rating comprises the following seven questions: 1) Was the exercise engaging?, 2) How fun was the activity?, 3) Did it feel like a physical exercise?, 4) Would you repeat performing this exercise?, 5) Would you perform it again in case of health necessity?, 6) Would you prefer this exercise over conventional training video that was shown in the YouTube video?, and 7) Would the audio-based feedback constitute an essential element in the exercise? Figure 2 shows the distribution of responses on questions 1 through 7 for Setting 1 and Setting 2.

\begin{figure}
\begin{center}
\subfigure[]{
\frame{\includegraphics[height=1.4in]{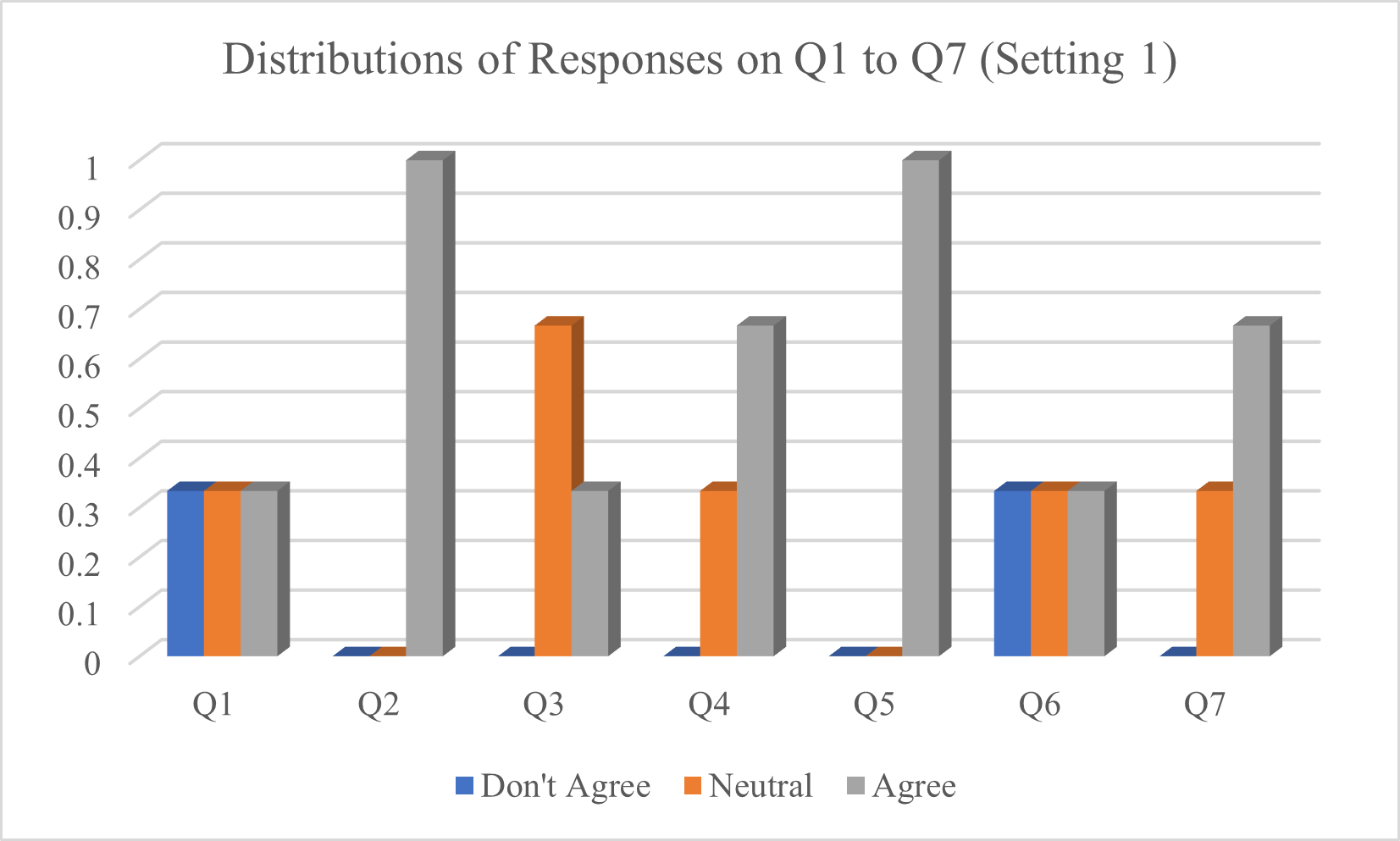}}}
\subfigure[]{
\frame{\includegraphics[height=1.4in]{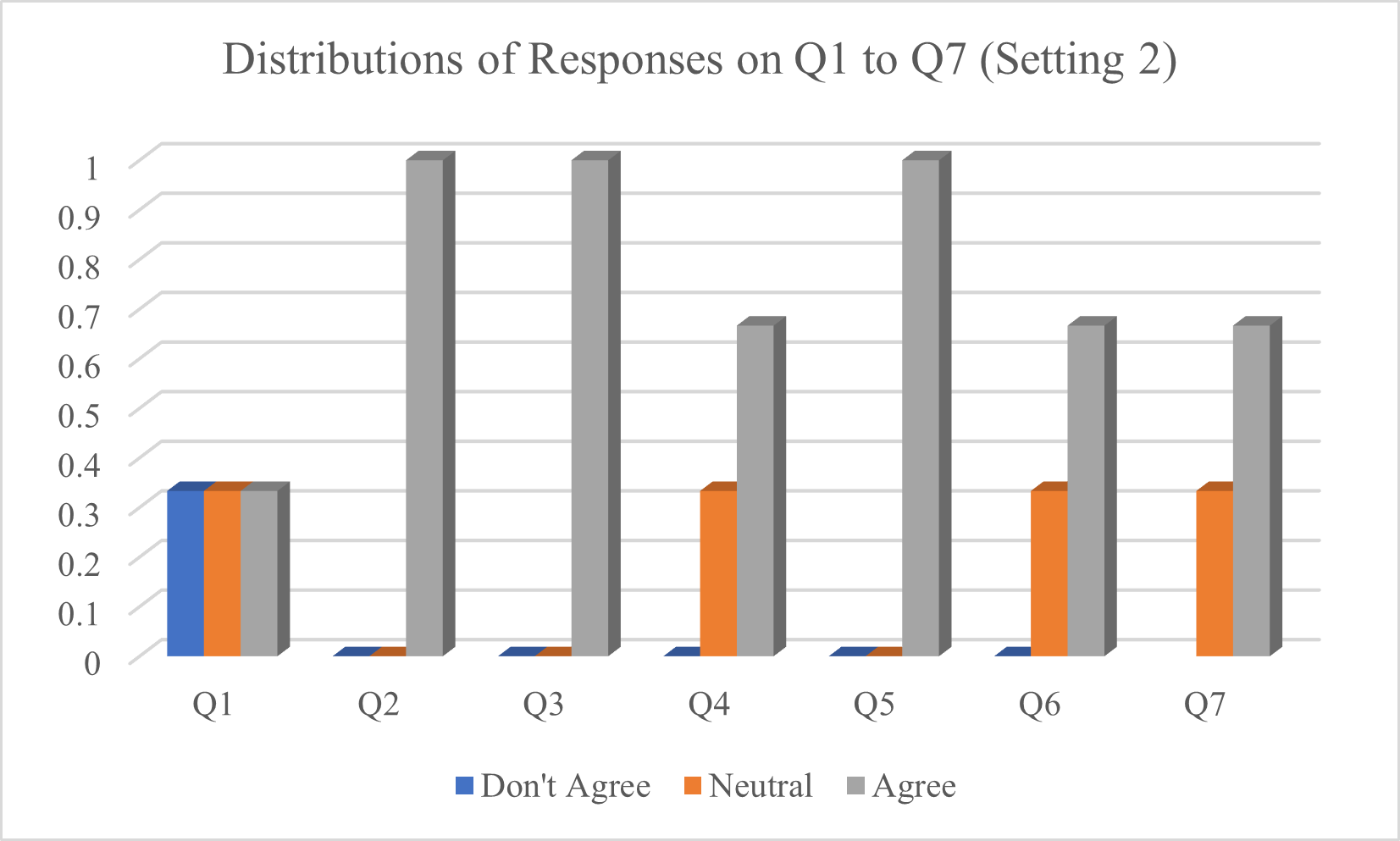}}}
\caption{Distribution of responses on question 1 to question 7--the subjective rating for engagement and preference of VR-Based intervention--for a) setting 1 and b) setting 2. The x-axis represents the seven questions representing the subjective rating.}
\label{Fig:DistResp}
\end{center}
\end{figure}

The responses of the participants show that although the exercise is engaging, they did not consider it fun. In this research, we adopt the following operational definitions for what is \textit{engaging} and what is \textit{fun}. An engaging activity is defined as a task that attracts a participant's attention to complete it, yet it does not get their interest. An activity is considered fun if a participant is interested in doing it, and hence it does not fail to acquire their attention by default. The participants might have found the exercise engaging but not interesting due to the nature and the environment of the task. Also, all of them practice e-sports regularly but only one of them goes to gym on a weekly basis. Towards better engagement, it might worth it to offer more diversity in the tasks required from the users. Also, sensory integration, such as auditory stimulation, might not be critical for strength exercises. However, as the training becomes more demanding, the influence of the sound element might become more apparent. This is something to be considered in future designs of the system. Moreover, the existence of a health urgency seems to be a significant factor for participants to use such a system, although they didn't consider the exercise to have a fun factor. Finally, the results shown in Fig.~\ref{Fig:DistResp} indicates that VR is still more appealing to the participants compared to traditional exercises. 

The subjective rating of the user experience comprises a set of seven aspects. These aspects are obstructive/supportive, complicated/easy, confusing/clear, boring/exciting, not interesting/interesting, conventional/inventive, and usual/leading edge. Table 1 shows the mean score for each of those aspects. For each aspect, the user either gives -1 or 1 if inclined towards one of the two sides, or zero if neutral.

\begin{table}[]
\begin{center}
\caption{The mean scores for each of the seven aspects included in the subjective rating of user experience. Please see text for more details.}
\begin{tabular}{|c|c|c|}
\hline
                         & \textbf{Mean Score} &                       \\ \hline
\textit{Obstructive}     & 1                   & \textit{Supportive}   \\ \hline
\textit{Complicated}     & 0.67                & \textit{Easy}         \\ \hline
\textit{Confusing}       & 0.33                & \textit{Clear}        \\ \hline
\textit{Boring}          & 0.33                & \textit{Exciting}     \\ \hline
\textit{Not interesting} & -0.67               & \textit{Interesting}  \\ \hline
\textit{Conventional}    & 0.67                & \textit{Inventive}    \\ \hline
\textit{Usual}           & 0                   & \textit{Leading Edge} \\ \hline
\end{tabular}
\end{center}
\label{tab:useexp}
\end{table}

The above table shows that further work should be done to make the instructions clearer to the user. The level of interesting-ness is expected in light of the previous subjective rating for engagement, since the content was found to be lacking fun by all the participants, although engaging. This results makes it more imperative to work on a larger variety of tasks. Moreover, although the setting might have seemed inventive, the potential of virtual reality as a visual stimuli still needs more improvement. This was apparent in the score of the last aspect which shows that the user were neutral, although the system design was meant to be leading edge. These mean scores will guide the development process of the future releases of the proposed system.

\section{Conclusion}
In this research, we studied the design and development of a VR-based training game that addresses neck pain, which we called Necknasium\textsuperscript{TM}. The system is comprised of a VR module that takes place in a gymnasium during which the user is instructed to do specific neck movements that are meant to help them manage their neck pain. We started by determining a set of training goals for both of the trainee and the therapist which were then translated into a set of design principles. %As a part of the implementation process, we shed light on a group of challenges that are related to tracking and quantifying neck movements using motion sensors. This was followed by discussing the alternative approach adopted in this study. 
Afterwards, we detailed a framework for developing this VR program/software and the various game elements that represent that framework. Towards evaluating the proposed system, we adopted a three-point scale for usability and user experience. On two undergraduate engineering students and a junior VR developer, we implemented this scale. With just three participants, this study provides preliminary insights on the feasibility of the proposed system. Based on the insights we acquired from the users' experience, we suggested a few directions for future enhancements of Necknasium\textsuperscript{TM}.

%\section*{Acknowledgements}
%The authors would like to thank the software development team of VRapeutic\textsuperscript{TM} who have contributed significantly to the system development process that made this research realizable. The authors would also like to thank the Information Technology Industry Development Agency (ITIDA) of Egypt who has funded this research (Project ID: CFP209).

\bibliographystyle{splncs04}
\bibliography{mybibliography}

%Author, F.: Article title. Journal \textbf{2}(5), 99--110 (2016)
%\end{thebibliography}
\end{document}